\documentclass[twocolumn]{autart}    
\usepackage{graphicx}         

\usepackage{amsmath,amsfonts,amssymb}
\usepackage{algorithmic}
\usepackage{algorithm}
\usepackage{array}
\usepackage[caption=false,font=footnotesize,labelfont=sf,textfont=sf]{subfig}
\usepackage{textcomp}
\usepackage{stfloats}
\usepackage{url}
\usepackage{verbatim}
\usepackage{graphicx}
\usepackage{cite}
\usepackage{tikz}
\usetikzlibrary{shapes,arrows, patterns}
\pgfdeclarepatternformonly{my north east lines}{\pgfqpoint{-2pt}{-2pt}}{\pgfqpoint{10pt}{10pt}}{\pgfqpoint{5pt}{5pt}}
{
  \pgfsetlinewidth{2pt}
  \pgfpathmoveto{\pgfqpoint{0pt}{0pt}}
  \pgfpathlineto{\pgfqpoint{9.1pt}{9.1pt}}
  \pgfusepath{stroke}
}
\usetikzlibrary{hobby}
\usepackage{enumerate}
\usepackage{xcolor}
\newtheorem{theorem}{Theorem}

\newtheorem{proposition}{Proposition}

\newtheorem{assumption}{Standing assumption}
\newtheorem{remark}{Remark}

\catcode`\@=11
\def\downparenfill{$\m@th\braceld\leaders\vrule\hfill\bracerd$}
\def\overparen#1{\mathop{\vbox{\ialign{##\crcr\crcr
\noalign{\kern0.4ex}
\downparenfill\crcr\noalign{\kern0.4ex\nointerlineskip}
$\hfil\displaystyle{#1}\hfil$\crcr}}}\limits}
\catcode`\@=12

\def\1{{\mathbbm 1}}   
\def\RR{{\mathbb R}}    
\def\cX{{\mathcal X}} 
\def\cD{{\mathcal D}} 
\def\cH{{\mathcal H}} 
 
\def\cV{{\mathcal V}} 
 
\def\cL{{\mathcal L}} 
\def\cF{{\mathcal F}} 
\def\cE{{\mathcal E}}

\usepackage{verbatim}

\DeclareMathOperator*{\argmin}{arg\,min}

\usepackage{textcomp}
\usepackage{multicol}

\newcommand{\dd}   {{\rm d}\hbox{\hskip 0.5pt}}

\newcommand{\rline}{{\mathbb R}}

\newcommand{\rfb}[1]{\mbox{\rm
   (\ref{#1})}\ifx\undefined\stillediting\else:\fbox{$#1$}\fi}

\newcommand\pdef[1]{\mathbb{S}_{\succ0}^{#1}}
\newcommand\psemidef[1]{\mathbb{S}_{\succeq0}^{#1}}
\newcommand\skewsymm[1]{\overline{\mathbb{S}}^{#1}}
\newcommand\symm[1]{{\mathbb{S}}^{#1}}
\def\bT{\mathbf{T}}

\newcommand\skewsymmfn[1]{\overline{\mathbf{S}}^{#1}}
\newcommand\symmfn[1]{{\mathbf{S}}^{#1}}

\begin{document}

\begin{frontmatter}
\title{Robust Neural IDA-PBC: passivity-based stabilization under approximations} 

\author[RUG]{Santiago Sanchez-Escalonilla Plaza}\ead{santiago.sanchez@rug.nl},    
\author[LAAS]{Samuele Zoboli}\ead{samuele.zoboli@laas.fr},               
\author[RUG]{Bayu Jayawardhana}\ead{b.jayawardhana@rug.nl}  

\address[RUG]{ENTEG-FSE, University of Groningen, Groningen, The Netherlands.}  
\address[LAAS]{LAAS-CNRS, Universit\'e de Toulouse, UPS, Toulouse, France.}             

\begin{keyword}                           
Port-Hamiltonian systems; Passivity-based control methods; IDA-PBC; Neural Networks; PINN.
\end{keyword}                            

\begin{abstract}                          
In this paper, we restructure the  Neural Interconnection and Damping Assignment - Passivity Based Control (\emph{Neural IDA-PBC}) design methodology, and we formally analyze its closed-loop properties.
Neural IDA-PBC redefines the IDA-PBC design approach as an optimization problem by building on the framework of Physics Informed Neural Networks (PINNs).
However, the closed-loop stability and robustness properties under Neural IDA-PBC remain unexplored. To address the issue, we study
the behavior of classical IDA-PBC under approximations. Our theoretical analysis allows deriving conditions for practical and asymptotic stability of the desired equilibrium point. Moreover, it extends the Neural IDA-PBC applicability to
port-Hamiltonian systems where the matching conditions cannot be solved exactly. Our renewed optimization-based design introduces three significant aspects: i) it involves a novel optimization objective including stability and robustness constraints issued from our theoretical analysis; ii) it employs separate Neural Networks (NNs), which can be structured to reduce the search space to relevant functions;
iii) it does not require knowledge about the port-Hamiltonian formulation of the system's model. Our methodology is validated with simulations on three standard benchmarks: a double pendulum, a nonlinear mass-spring-damper and a cartpole. Notably, classical IDA-PBC designs cannot be analytically derived for the latter. 
\end{abstract}

\end{frontmatter}

\section{Introduction}
Interconnection and Damping Assignment Passivity Based Control (IDA-PBC) proved to be very effective in controlling port-Hamiltonian systems \cite{robustacosta2014,IDAPBCuav,IDAPBCbeam}. Its success is due to its ability to readily define the closed-loop system's behavior \cite{ortega2002IDA}. However, its design may turn out to be challenging, as it requires the solution of a set of (nonlinear) partial differential equations (PDEs) known as \emph{matching equations}. Often, deriving closed-form solutions for the matching equations requires several simplification steps. These include nontrivial changes of coordinates \cite{changeofcoordinates2007}; imposing strict assumptions on the system dynamics \cite{acosta2004IDA}; reducing parameters through the assignment of desired local dynamics \cite{kotyczka2013local}; or transforming the PDEs into algebraic inequalities \cite{acosta2009ineq}. While simplifications can facilitate controller synthesis, they often necessitate compromises in other metrics that might be crucial in performance-critical applications (see, e.g., \cite{robustacosta2014}). For a detailed exploration of the challenges inherent in this type of control synthesis method, including its ill-posed nature, see \cite{ortega2004survey,nageshrao2016survey}.

These drawbacks motivate the need for a methodology that enables the synthesis of PBC methods without solving nonlinear PDEs analytically, while preserving the mathematical integrity essential for analyzing the closed-loop system. Such an approach would enhance the applicability of energy-based control design methods in practical scenarios, connecting theoretical and real-world implementations.
A promising direction involves leveraging numerical approximations to find solutions to nonlinear PDEs. Among various methods, Physics-Informed Neural Networks (PINNs)
\cite{raissi2019PINN} have become the de-facto approach for obtaining the approximate solutions \cite{Cuomo2022ScientificML,antonelo2022physicsinformed,GOKHALE2022118852,NICODEMUS2022331,Yuanyuan2022}. Their popularity is due to the fact that neural networks (NNs) are known as universal approximators \cite{hornik1989universal,cybenko1989universal} and do not require specific discretization schemes thanks to Automatic Differentiation tools \cite{baydin2015Adifferentiation}. Moreover, they can help overcome the curse of dimensionality \cite{han2017curse}. 

Automatic discovery of IDA-PBC solutions using PINNs has been explored as \emph{Neural IDA-PBC} in \cite{sanchez2021total} and \cite{sanchez2022under}. \emph{Neural IDA-PBC} presents a methodology that streamlines the approximation of solutions to the matching equations, eliminating the necessity for prior transformations or assumptions related to the underactuated coordinate's dependency \cite{acosta2004IDA}. This advancement not only simplifies the design process, but also expands the applicability of IDA-PBC to nonacademic scenarios.
However, the effect of the numerical approximation on the closed-loop behavior remains unexplored, thus hindering its relevance to real-world problems.
In this paper, we address this issue by studying and analyzing the effect of approximation/optimization errors on the stability and robustness properties of the closed-loop system, providing guarantees on the asymptotic behavior. Furthermore, we improve on the interpretability of \emph{Neural IDA-PBC} by using independent NNs to parameterize the missing functions in IDA-PBC. Our modular representation of the NNs additionally allows embedding different properties in the construction of the NN functions, in turn guaranteeing \emph{desirable} structural constraints (e.g. skew-symmetry, positivity, etc.) by design. Moreover, our approach also shows the feasibility of a post-training \emph{tunable} parameter to modify the closed-loop transient behavior. This feature is particularly advantageous in situations with performance requirements. 
Given the known complexity in selecting the optimal hyperparameters for the NNs--such as the network architecture or sampling strategy--our methodology operates under the assumption that a solution to the stabilization problem is within reach and can be discovered using conventional optimization algorithms.

The paper is structured as follows. Section II introduces preliminary notions in IDA-PBC, revisiting a pivotal result on robust IDA-PBC first established in \cite{becherif2005robustIDA}. This section sets the stage for our contributions by outlining the theoretical underpinnings of our work. Section III explores how mismatches in the IDA-PBC matching equations affect the closed-loop port-Hamiltonian system under the \emph{Neural IDA-PBC} control law. We provide a detailed analysis proving that the desired equilibrium point $x^\star$ can be asymptotically stabilized under assumptions on the approximation quality of the matching equation. Then, we discuss the relaxation of these assumptions, which leads to the asymptotic stabilization of an equilibrium point in the neighborhood of $x^\star$. The distance between the two points is related to the NN approximation errors. Section IV delves into the numerical implementation of our \emph{Neural IDA-PBC} methods. Using mathematical insights on port-Hamiltonian systems, including the specific structures of systems' matrices, the Hamiltonian functions, and the energy dissipation property, we detail the structure of each NN designed to approximate functions required by IDA-PBC, including the formulation of specific loss functions.
Finally, Section V showcases the efficacy of our methods through numerical simulations, affirming the theoretical analyses and implementation strategies discussed in previous sections.

\section{Preliminaries}

{\bf Notations.} 
We denote by $\rline_{>0}, \rline_{\ge0}$ the sets of strictly positive and non-negative real numbers, respectively. We identify by $\pdef{n},\psemidef{n}, \symm{n}, \skewsymm{n}$ the sets of positive definite, positive semi-definite, symmetric, and skew-symmetric real matrices of dimension $n\times n$. The mapping $\symmfn{n}:\mathbb{R}^{\frac{n(n+1)}{2}}\to \symm{n}$ is the symmetric matrix reconstruction map generating an $n \times n$ symmetric matrix from $\frac{n(n+1)}{2}$ independent elements. The mapping $\skewsymmfn{n}: \mathbb{R}^{\frac{n(n-1)}{2}}\to \skewsymm{n}$ is the skew-symmetric matrix reconstruction map generating an $n \times n$ skew-symmetric matrix from $\frac{n(n-1)}{2}$ independent elements. For square matrices $A,B$, the notation $A\preceq B$ stands for $(A-B)\preceq 0$ and $\sigma_{\min}(A)$ denotes the eigenvalue of $A$ with smallest real part. 
For a given matrix-valued function $g:\rline^n\to \rline^{n\times m}$, we denote by $g^\perp:\rline^n\to \rline^{(n-m)\times n }$ its left annihilator satisfying $g^\perp(x)g(x)=0$. We use $\|\cdot\|$ as the norm operator for matrices and vectors. For a given set $\cX$, we identify its boundary by $\partial \cX$. The notation $\cX \setminus \mathcal G$ identifies the set difference operator between $\cX$ and $\mathcal G$.
When a set $\cX$ is strictly included in a set $\mathcal G$, we use $\cX \subsetneq \mathcal G$. Otherwise, we use $\cX \subseteq \mathcal G$. 
We denote the compact sub-level set of a \textit{proper} function $H:\RR^n\to\RR_{\ge0}$ parametrized by $\varepsilon>0$ with $\cH_\varepsilon = \{x\in\cX:H(x)\le \varepsilon\}$ and its boundary by $\partial \cH_\varepsilon$. 

\subsection{Port-controlled Hamiltonian systems}

Throughout the paper, we consider port-controlled Hamiltonian systems, as described in \cite{schaft2014PH}. Hence, we adopt an energy-based viewpoint to describe the dynamics of multi-domain physical systems via internal and external power flows. 
The evolution of port-Hamiltonian systems with dissipation is described by the dynamics
\begin{equation}
\begin{array}{rl}
&\left.\begin{array}{rl}\dot{x} & = f(x)+g(x)u \\
y & = g^\top(x)\dfrac{\partial H}{\partial x}(x)
\end{array}\right\},\\
&f(x):= \Big( J(x)-R(x) \Big) \dfrac{\partial H}{\partial x}(x), 
\label{eq:OLPH}
\end{array}
\end{equation}
where $x\in\RR^n$ is the state, $u,y\in\RR^m$ are the input and output power-conjugate variables, $J:\RR^n\to\skewsymm{n}$ is the skew-symmetric interconnection matrix-valued function corresponding to the internal power-conserving structure of physical systems, $R:\RR^n\to\psemidef{n}$ is a positive semi-definite damping matrix-valued function that accounts for energy losses, $H:\RR^n\to\RR_{\ge0}$ is a two-times differentiable Hamiltonian function  (i.e., the total energy stored in the system) and $g:\RR^n\to\RR^{n\times m}$ is the matrix-valued function that  maps the input variable to the system dynamics. Following classical assumptions in the port-Hamiltonian framework, we assume that $m\leq n$ and that all the above functions are sufficiently smooth to ensure forward completeness of the trajectories of the autonomous system. The Hamiltonian function $H$ is assumed to be {\it proper}, i.e., all sublevel sets are compact, or it is radially unbounded.

The focus on power exchange as a proxy for modeling system dynamics allows for a very powerful framework that simplifies the analysis and design of stabilizing controllers. PBC aims at designing a feedback control law $u = \beta(x)$ such that the closed-loop system satisfies the following \emph{passivation/dissipation} objective
\begin{equation}
    H_d\Big(x(t)\Big) - H_d\Big(x(0)\Big) \leq \int_0^t -\dfrac{\partial H_d}{\partial x}^\top R_d(x(\tau))\dfrac{\partial H_d}{\partial x}\dd \tau,
    \label{eq:passivation}
\end{equation}
where $H_d:\rline^{n}\to\rline_{\ge0}$ is the desired energy function, which is (locally) strictly convex and has a minimum at the desired state $x^\star$, and $R_d:\RR^n\to\psemidef{n}$ is the desired dissipation function. 
The \emph{passivation/dissipation} objective \eqref{eq:passivation} implies that there exists a neighborhood of $x^\star$ that is \textit{forward-invariant}, namely, that all system trajectories starting in this neighborhood remain in it.

\subsection{Interconnection and Damping Assignment-PBC}\label{sec:IDA-PBC}
IDA-PBC is a PBC control scheme that achieves stable closed-loop dynamics around an equilibrium $x^\star \in \cX \subseteq \RR^n$ by designing a control law $u=\beta(x)$ that satisfies the \emph{passivation/dissipation} objective \eqref{eq:passivation} and ``shapes'' the closed loop dynamics into a desired port-Hamiltonian system 
\begin{equation}
\left.\begin{array}{rl}\dot{x} & = f_d(x) = \Big( J_d(x)-R_d(x) \Big) \dfrac{\partial H_d}{\partial x}(x)\\
y & = g^\top(x)\dfrac{\partial H_d}{\partial x}(x)
\end{array}\right\},
\label{eq:CLPH}
\end{equation}
with $J_d: \cX \rightarrow \skewsymm{n}$, $R_d: \cX \rightarrow \pdef{n}$ and $H_d: \cX \rightarrow \RR_{\ge0}$, the desired interconnection matrix, damping matrix and energy function of the closed-loop system  \cite{ortega2002IDA}.
Asymptotic stability of $x^\star$ is guaranteed if the storage function $H_d(x)$ satisfies the following conditions:
\begin{equation}
    \begin{array}{rl}
        \frac{\partial^2 H_d}{\partial x^2}(x)& \succ 0 \quad \forall x\in \cX,\\
        H_d(x)&>0 \quad \forall x\in \cX \setminus \{x^\star\},\\
        H_d(x^\star) & = 0, \quad \frac{\partial H_d}{\partial x}(x^\star) =0.
    \end{array}
\label{eq:hamiltonian}
\end{equation}

We now recall a controller design achieving stabilization and shaping the closed-loop.

\begin{proposition}[{\cite[Proposition 1]{ortega2004survey}}]
\label{proposition:IDAPBC}
Consider the port-Hamiltonian system \eqref{eq:OLPH}, suppose $H_d$ satisfies \eqref{eq:hamiltonian} 
and assume the triplet $(H_d,J_d,R_d)$ 
is a solution to the matching equation
\begin{equation} \label{eq:match}
    g^{\perp}(x)[f(x)-f_d(x)]=0.
\end{equation}
Let $u=\beta(x)$ with
\begin{align}
    \beta(x) := [g^\top(x) g(x)]^{-1}g^\top(x)\Big(f_d(x)-f(x)\Big).
    \label{eq:u}
\end{align}
Then, the dynamics of the closed-loop \eqref{eq:OLPH} and \eqref{eq:u} are equivalent\footnote{Namely, for any initial condition $x^\circ\in\cX$ the corresponding trajectories $X(t,x^\circ), X_d(t,x^\circ)$ generated by systems  \eqref{eq:OLPH},\eqref{eq:u} and \eqref{eq:CLPH}, respectively, satisfy $X(t,x^\circ)= X_d(t,x^\circ)$ for all $t\ge0$.} to \eqref{eq:CLPH} and $x^\star = \arg \min H_d(x)\in \cX$ is a stable equilibrium point. Furthermore, if
$R_d(x)\in\pdef{n}$ for all $x\in\cX$,
the equilibrium point $x^\star$ is asymptotically stable with domain of attraction $\cX$. 
\label{proposition:idapbc}
\end{proposition}\vspace{0.2cm}

\subsection{Robust IDA-PBC}\label{sec:Robust-IDA-PBC}
The major limitation of IDA-PBC arises from the solvability of the 
\emph{matching equations}  \eqref{eq:match}. In other words, finding a suitable combination of $H_d, J_d$ and  $R_d$ to define the control law \eqref{eq:u} presents a challenging task. 
To address the issue, multiple approaches have been proposed in the literature and we refer interested readers to \cite{ortega2004survey} for a comprehensive discussion. 

More specifically, finding an \textit{exact} solution to \eqref{eq:match} requires \textit{exact} model matching of the open-loop dynamics,
which often limits the application of the IDA-PBC controller to
theoretical exercises. Therefore, relaxing this requirement is essential to extend the applicability of this technique. 
One way to relax this condition is to include a small perturbation in the target dynamics \eqref{eq:CLPH}, which can be used as an additional degree of freedom during the design of the IDA-PBC controller. Moreover, this addition makes the closed-loop system robust to affine disturbances as in \cite{becherif2005robustIDA}. 
In what follows, we recall a stability result for IDA-PBC with perturbed desired dynamics. We also present the initial steps of the proof, as they will be useful for our analysis of IDA-PBC under approximate matching conditions.

\begin{theorem}[{\cite[Theorem 1]{becherif2005robustIDA}}]
\label{theorem:robustIDAPBC}
Consider system \eqref{eq:CLPH} and assume there exists a scalar $r>0$ such that 
$R_d(x)\succeq r I$ for all $x\in\cX$. Let $\xi:\RR^n\to\RR^n$ be a bounded function such that 
\begin{equation}
\|\xi(x)\| \leq \kappa \left\|\dfrac{\partial H_d}{\partial x}(x)\right\| , \qquad \forall x\in \mathcal X,
\label{eq:robust}
\end{equation}
 for some $\kappa\in(0,r)$. Then $x^\star= \arg \min H_d(x)\in \cX$ is an asymptotically stable equilibrium with 
 domain of attraction $\cX$ for  the  port-Hamiltonian system
\begin{equation}
\left.\begin{array}{rl}\dot{x} & = 
f_d(x) + \xi(x)\\
y & = g^\top(x)\dfrac{\partial H_d}{\partial x}(x).
\end{array}\right\}
\label{eq:CLPHxi}
\end{equation}
\end{theorem}

\begin{proof}
Let $H_d$ be a candidate Lyapunov function for \eqref{eq:CLPHxi}. 
Due to the positive definiteness of $R_d(x)$ and condition \eqref{eq:robust}, the computation of its Lie derivative yields
\begin{align}
\begin{split}
        \dot{H}_d(x)& \leq - \left(r-\kappa\right)\left\|\dfrac{\partial H_d}{\partial x}(x)\right\|^2 <0.
        \label{eq:dhdt}
\end{split}
\end{align}
The rest of the proof follows vis-\`a-vis the one of \cite[Proposition 1 ]{ortega2004survey}.  
\end{proof}
According to Theorem~\ref{theorem:robustIDAPBC},  (local) asymptotic stability of $x^\star$ is guaranteed by control law \eqref{eq:u} even in the presence of  additive disturbance, provided that it is vanishing at the desired equilibrium point $x^\star$.

\section{Neural IDA-PBC}
As presented in Section II,
the family of IDA-PBC controllers capable of stabilizing 
\eqref{eq:OLPH} around the desired equilibrium $x^\star$ can be formulated as an inverse problem that consists of finding the tuple of functions $\bT:=(g^\perp, J_d, R_d, H_d)$ satisfying \eqref{eq:hamiltonian} and \eqref{eq:match}. Developing an {\it ad hoc} method to choose the right combination of functions that achieve the desired closed-loop systems requirements is, in general, an intractable task. Correspondingly, systematic procedures such as the Non-parameterized IDA, Algebraic IDA or Parameterized IDA, have been proposed to reduce the function search space, resulting in amenable results \cite{ortega2004survey}. 
The objective of this section is to formalize and restructure the machine-learning 
design framework proposed in \cite{sanchez2021total,sanchez2022under}. Hence, we encode the IDA-PBC desired dynamic selection  into the optimization problem 
\begin{equation}
    \bT^\star= \argmin_{\bT\in\cF^4}\left\| g^\perp(x)[f(x)-f_d(x)]\right\|
   \label{eqn:neuralIDAopt}
\end{equation}
where $\cF^4$ is the function space of universal approximators that encapsulate the locally Lipschitz functions $g^\perp, J_d, R_d$ and the twice continuously differentiable function $H_d$. 
Hence, to avoid ill-conditioned optimization problems, throughout the rest of the paper we will assume the following
\begin{assumption}\label{as:approx_match}
     For any small scalar $\epsilon> 0$, there exists a solution $\bT_\epsilon=(g_\epsilon^\perp, J_{d,\epsilon}, R_{d,\epsilon}, H_{d,\epsilon})$ such that  
\[
\left\| g_\epsilon^{\perp}(x)\Big[f(x)-(J_{d,\epsilon}(x)-R_{d, \epsilon}(x))\frac{\partial H_{d,\epsilon}}{\partial x}(x)\Big]\right\| \le \epsilon, 
\]
 for all $x\in\cX$.
\end{assumption}

Our first objective 
is then to understand the influence of approximation errors in the solution of \eqref{eqn:neuralIDAopt} onto the port-Hamiltonian structure of the closed-loop systems. In particular, we focus on approximations of the functions\footnote{The IDA-PBC method offers the option of using $g^\perp$ as an additional degree of freedom for solving the matching-equation. Yet, considering the simplicity of the requirement it imposes, specifically $g^\perp g = 0$, we choose not to look for an approximation of this function. Instead, we opt to model an exact left-annihilator in each case. While constraining the space of solutions, this choice simplifies the optimization problem by reducing the number of functions that need to be found.} $J_d,R_d$ and $H_d$.
The effect of approximation is formally presented in the following proposition\footnote{Note that we could group the effect of uncertain dynamics with the approximation error. However, due to the similarity in the analysis and the limited scope of this paper, we choose to solely focus on approximation errors.}. 
\begin{proposition}
\label{prop:externaldisturbance}
    Let $\widetilde \bT = (\Tilde{J}_d, \Tilde{R}_d, \Tilde{H}_d)$ be a tuple of function approximations such that Standing assumption~\ref{as:approx_match} holds for some $\epsilon>0$ with $\bT_\epsilon = (g^\perp(x), \widetilde \bT)$.  Moreover,
    suppose that $g^\perp(x)$ spans the kernel space of $g(x)$ for all $x\in\cX$. 
    Let $u=\Tilde{\beta}(x)$ with
\begin{equation}
    \begin{array}{rl}
    \Tilde{\beta}(x) &:= [g^\top(x) g(x)]^{-1}g^\top(x)\Big(\Tilde{f}_d(x)-f(x)\Big),\\
    \Tilde{f}_d(x)&:= [\tilde J_d(x) - \tilde R_d(x)]\dfrac{\partial \tilde H_d}{\partial x}(x).
    \end{array}
    \label{eq:tilde_u}
\end{equation}
    Then, 
    the closed-loop port-Hamiltonian system \eqref{eq:OLPH}, \eqref{eq:tilde_u} is equivalent\footnotemark[1] to
    \begin{equation}\label{eq:CLPH_approx}
        \dot x = \tilde f_d(x) +\xi(x),
    \end{equation}
    where $\xi:\RR^n\to\RR^n$ is an affine state-dependent perturbation satisfying
   \begin{align} \label{eqn:bounded_mismatch}
       \|\xi(x)\| \leq
       \|G(x)^{-1}\| \, \|\mu(x)\|, \quad \forall x\in \cX,
   \end{align}
   with $G(x):=\begin{pmatrix}
g^\top(x)\\ g^\perp(x)
\end{pmatrix}
\in\RR^{n\times n}$ a full-rank matrix for all $x\in\cX$ and $\mu$ the \emph{mismatch} function 
   \begin{equation}
    \mu(x):= -g^\perp(x)[f(x) -\tilde f_d(x)]. \label{eqn:mismatch}
    \end{equation}
\end{proposition}

\begin{proof}
Pre-multiplying \eqref{eq:OLPH} by $G(x)$, we obtain
\begin{equation*}
    G(x)\dot x = \begin{pmatrix}
    g(x)^\top f(x)\\
    g(x)^\perp f(x)
    \end{pmatrix} +
    \begin{pmatrix}
    g(x)^\top g(x) u\\
    0
    \end{pmatrix}.
\end{equation*}
By the choice of the control law $u=\tilde \beta(x)$, \eqref{eq:tilde_u} and \eqref{eqn:mismatch}, it follows from the above equation that 
\begin{equation*}
    G(x)\dot x = \begin{pmatrix}
    g(x)^\top \tilde f_d(x)\\
    g(x)^\perp \tilde f_d(x)
    \end{pmatrix} +
    \begin{pmatrix}
    0\\
    \mu(x)
    \end{pmatrix}.
\end{equation*}
Finally, by defining $\xi^\top(x) := (0 \quad \mu(x))^\top (G(x)^{-1})^\top$ and by pre-multiplying both sides by the inverse of $G(x)$, we arrive at \eqref{eq:CLPH_approx}. 
Moreover, by the definition of $\xi$ and \eqref{eqn:mismatch}, inequality \eqref{eqn:bounded_mismatch} holds, thus concluding the proof.
\end{proof}\vspace{0.2cm}

Proposition~\ref{prop:externaldisturbance} shows the effect of approximating the desired function $f_d$. Following standard IDA-PBC theory, the closed-loop dynamics will be equivalent to the learned function $\tilde f_d$ with an additive disturbance generated by  
a nonzero mismatch $\mu$ in \eqref{eqn:mismatch}. Observe that \eqref{eq:CLPH_approx} closely resembles \eqref{eq:CLPHxi}. 
Hence, we can study the influence of a \emph{mismatch} in the matching equations by exploiting techniques similar to the ones in Theorem~\ref{theorem:robustIDAPBC}. However, differently from \eqref{eq:robust}, $\xi$ in \eqref{eqn:bounded_mismatch} may not be vanishing in the equilibrium point as $\mu$ can be non-singular in $x^\star$. This can lead to an ill-posed condition on the desired damping function $\tilde R_d$. Indeed, as shown by inequality \eqref{eq:dhdt} in the proof of Theorem \ref{theorem:robustIDAPBC}, the time-derivative of $H_d$ is negative if 
\begin{align*}
\left\|G^{-1}(x)\right\| \|\mu(x)\| <  r \left\|\dfrac{\partial \tilde H_d}{\partial x}(x)\right\| 
, \quad \forall x\in\cX,
\end{align*}
where $r$ can be interpreted as
the eigenvalue of $\tilde R_d(x)$ with smallest real part. 
Hence, a necessary (yet not sufficient) condition for the above inequality to hold is 
\begin{equation*}
    \sigma_{\min}(\tilde R_d(x)) > \left\|G^{-1}(x)\right\| \inf_{x\in\cX} \|\mu(x)\|\left\|\dfrac{\partial \tilde H_d}{\partial x}(x) \right\|^{-1}. 
\end{equation*} 

By the structural properties \eqref{eq:hamiltonian}, for any $x\in \cX$ we have $\lim_{s\to 0}\left\|\frac{\partial \tilde H_d}{\partial x}(x^\star + sx)\right\| =0.$ Thus, if there exists a constant $\underline \mu>0$ such that $\|\mu(x)\|\ge \underline \mu$ for all $x\in\cX$, the fulfillment of the above inequality implies $\sigma_{\min}(\tilde R_d(x))\to\infty$ as $x\to x^\star$, since $\left\|G^{-1}(x)\right\|$ and the infimum take finite non-zero values. However, if there exists a bounded constant $\bar \kappa \ge 0$ such that 
\begin{equation}\label{eq:mu_kappa_bound}
\lim_{s\to 0}\|\mu(x^\star+ sx)\|\left\|\dfrac{\partial \tilde H_d}{\partial x}(x^\star+ sx) \right\|^{-1}=\bar \kappa,
\end{equation}
we can avoid ill-posed constraints on $\tilde R_d(x)$.
Therefore,
we now present a result showing that, under assumptions akin to \eqref{eq:mu_kappa_bound}, stabilization of the port-Hamiltonian system \eqref{eq:OLPH} at the desired equilibrium point $x^\star$ is achievable with a neural network-based approximation $(\Tilde{J}_d, \Tilde{R}_d, \Tilde{H}_d)$ with locally Lipschitz $\Tilde{J}_d, \Tilde{R}_d$ and twice continuously differentiable $\Tilde{H}_d$.
\begin{proposition}\label{prop:vanishing_mu}
Consider the system \eqref{eq:OLPH} and suppose that the neural networks $\Tilde{J}_d, \Tilde{R}_d, \Tilde{H}_d$ satisfy the following structural conditions:
\begin{enumerate}[(i)]
    \item For all $x\in\cX$,
    \[
    \Tilde{J}_d(x) \in \skewsymm{n}, \quad \Tilde{R}_d(x) \in \pdef{n}, \quad 
        \frac{\partial^2 \tilde H_d}{\partial x^2}(x) \succ 0.
    \]
    \item $\tilde H_d(x)>0$ for all $x\in \cX \setminus \{x^\star\}$.
    \item $\tilde H_d(x^\star)  = 0, \quad \dfrac{\partial \tilde H_d}{\partial x}(x^\star)=0 .$
\end{enumerate}
     
Moreover, suppose  that $\mu$ in \eqref{eqn:mismatch} satisfies \begin{equation}
\label{eqn:bounded_mu_bar}
    \|\mu(x)\| \leq {\kappa}{\|G(x)\|}\left\|\dfrac{\partial \tilde H_d}{\partial x}(x)\right\|, \quad \forall x\in\cX, 
    \end{equation}
    with $\kappa\in(0,\sigma_{\min}(\tilde R_d(x)))$ for all $x\in\cX$.
    Then the \emph{Neural IDA-PBC} control law $u=\tilde \beta(x)$ in \eqref{eq:tilde_u} with $\Tilde{J}_d, \Tilde{R}_d$, and $\Tilde{H}_d$ asymptotically stabilizes $x^\star$. 

\end{proposition}
\begin{proof}
   The proof follows directly from equation \eqref{eqn:bounded_mismatch} in Proposition \ref{prop:externaldisturbance} and Theorem \ref{theorem:robustIDAPBC}.
\end{proof}

In general, the assumption of vanishing $\mu(x)$ in \eqref{eqn:bounded_mu_bar} as considered in Theorem \ref{prop:vanishing_mu} can cover port-Hamiltonian systems that admit a local exact solution to the IDA-PBC matching equation \eqref{eq:match} with a (strictly) positive-definite dissipation $R_d(x)$. In this case, by ensuring that the function approximators $\Tilde{J}_d, \Tilde{R}_d, \Tilde{H}_d$ match ``\emph{exactly}'' the functions $J_d, R_d, H_d$ solution of \eqref{eq:match} in $x^\star$ (e.g., with blending techniques similar to \cite{zoboli2021reinforcement}), the mismatch caused by the approximation $\mu(x)$ will satisfy \eqref{eq:mu_kappa_bound}.
However, the result of Proposition \ref{prop:vanishing_mu} is not applicable if the corresponding port-Hamiltonian system does not admit a solvable IDA-PBC matching equation to stabilize $x^\star$ with positive definite $\tilde R_d(x)$, or if the existence of an exact solution to \eqref{eq:match} is not ensured (see Standing assumption~\ref{as:approx_match}). Typical examples of such port-Hamiltonian systems are under-actuated mechanical systems in the presence of dissipation \cite{GOMEZESTERN2004451}.
In this context, the requirement of vanishing $\mu(x)$ becomes the hardest constraint to satisfy. In fact, the sole assumption of upper-boundedness of the mismatch (i.e., $\|\mu(x)\| \leq \bar\mu$ for some $\bar\mu>0$) can impose an infeasible condition on the desired damping matrix $\tilde R_d(x)$.

Hence, we now aim to analyze the influence of a non-vanishing mismatch $\mu(x)$ coupled with a bounded damping matrix $\tilde R_d(x)$ on the stability of the closed-loop systems.
In the following, we will consider the case where $\mu$ is locally Lipschitz and \eqref{eq:robust} only holds outside a compact set containing $x^\star$. In other words, we look for asymptotic properties of the perturbed closed-loop system \eqref{eq:CLPH_approx} when \eqref{eq:robust} does not hold in a neighborhood of $x^\star$. To this aim, consider an $\varepsilon$-parametrized compact\footnote{The compactness of $\cH_\varepsilon$ is due to the assumption of proper $H$, after \eqref{eq:OLPH}.} sub-level set of $\tilde H_d$, denoted $\cH_\varepsilon$. 
Moreover, define the \emph{dissipation set} $\cD:=\cX\setminus\cH_\varepsilon$. 
We have the following result on the practical stabilization of the closed-loop port-Hamiltonian system under non-vanishing perturbation $\xi(x)$ and bounded dissipation matrix $\tilde R_d(x)$.   

\begin{proposition}
\label{prop:match_ineq_and_dissip_relax}
Consider the port-Hamiltonian system with affine disturbance \eqref{eq:CLPHxi} and assume there exists $\varepsilon>0$ such that
\begin{equation}
    \dfrac{\partial \tilde H_d^\top}{\partial x}(x) \tilde R_d(x)\dfrac{\partial \tilde H_d}{\partial x}(x) > \dfrac{\partial \tilde H_d^\top}{\partial x}(x)\xi(x)
    \label{eq:damping_assign_relax}
\end{equation}
holds for all $x\in \cD\, \cup\, \partial \cH_\varepsilon$. Then $\cH_\varepsilon$ is attractive and forward invariant. Moreover, system \eqref{eq:CLPHxi} has an equilibrium $\bar x \in \cH_\varepsilon$ satisfying 
\begin{equation}
 \Big(\tilde J_d(\bar x)-\tilde R_d(\bar x)\Big)\dfrac{\partial \tilde H_d}{\partial x}(\bar x) = - \xi(\bar x).
 \label{eq:fixedpoint}
\end{equation}
\end{proposition}

\begin{proof}
Similar to the proof of Theorem \ref{theorem:robustIDAPBC}, we obtain ${\dot {\tilde H}_d(x)<0}$ for all $x\in\cD$. This implies that $\cH_\varepsilon$ is attractive and forward invariant. Moreover, being $\tilde H_d$ a valid Lyapunov function up to $\cH_\varepsilon$, $\partial \cH_\varepsilon$ is homeomorphic to a unit sphere, \cite[Corollary 2.3]{wilson1967structure}\footnote{Thanks to Freedman \cite{freedman1982topology}  and Perelman \cite{morgan2007ricci} the restriction on dimensions is not needed.}. Hence, by \cite[Chapter I, Theorem 8.2]{hale1969ordinary}, there exists a fixed point $\bar x \in \cH_\varepsilon$ satisfying \eqref{eq:fixedpoint},
and this concludes the proof.
\end{proof}

Proposition \ref{prop:match_ineq_and_dissip_relax} ensures the existence of a shifted equilibrium $\bar x\in \cH_\varepsilon$ with practical stability properties for system \eqref{eq:CLPHxi} when \eqref{eq:robust} is not satisfied everywhere. Moreover, it ensures that $\bar x$ lies in a neighborhood of $x^\star$, as both belong to $\cH_\varepsilon$. Following our discussion on the infeasibility of the constraints on $\tilde R_d$ when $\xi$ is non-vanishing, the smaller the size of $\cH_\varepsilon$, the stronger this constraint \eqref{eq:robust} is on $\tilde R_d(x)$. Indeed, reducing $\cH_\varepsilon$ will lead to $\bar x$ being closer to $x^\star$. Differently put, since $\bar x, x^\star \in \cH_\varepsilon$ for all $\varepsilon \ge 0$, by reducing $\varepsilon$ the distance between the two points decreases, as $\cH_\varepsilon$ shrinks due to the convexity of $\tilde H_d$.

Nonetheless, even for a non-vanishing $\mu$, we can establish asymptotic properties for the closed-loop. In the following theorem, we prove asymptotic stability of the above-mentioned shifted equilibrium point $\bar x$  when the function approximators $\Tilde{J}_d, \Tilde{R}_d, \Tilde{H}_d$ provide a sufficiently good approximation.

\begin{theorem}\label{thm:asym_stab_shifted_eq}
Consider system \eqref{eq:OLPH}, assume the neural networks $\Tilde{J}_d, \Tilde{R}_d, \Tilde{H}_d$ satisfy the structural constraints (i)-(iii) in Proposition~\ref{prop:vanishing_mu}, and let \eqref{eqn:mismatch} hold with a continuously differentiable bounded function $\mu$ satisfying $\|\mu(x)\|\leq \bar \mu$, for some $\bar \mu>0$. Let $\cH_\varepsilon\subsetneq \cX$ be a sub-level set\footnotemark[3] of $\tilde H_d$.
Then, for any $\varepsilon>0$ such that
\begin{equation}
    \dfrac{\partial \tilde H_d^\top}{\partial x}(x) \tilde{R}_d(x)\dfrac{\partial \tilde H_d}{\partial x}(x) > \dfrac{\partial \tilde H_d^\top}{\partial x}(x)G(x)^{-1}\!\!\begin{pmatrix}0 \\ \mu(x)\end{pmatrix}
    \label{eq:damping_assign_relax_tilde}
\end{equation}
holds for all $x\in \cD \, \cup\, \partial \cH_\varepsilon$, with $G(x)$ as in Proposition~\ref{prop:externaldisturbance}, there exists $\bar \mu^\star>0$ such that, if
\begin{equation}\label{eqn:mu_bound_les}
\|\mu(x)\|\leq \bar \mu^\star, \quad \left\|\frac{\partial \mu}{\partial x}(x)\right\|\leq \bar \mu^\star,  \quad \forall x\in \cH_\varepsilon,
\end{equation} the 
closed-loop \eqref{eq:OLPH},\eqref{eq:tilde_u} has a locally asymptotically stable equilibrium point $\bar x\in \cH_\varepsilon$.
\end{theorem}

\begin{proof}
The proof is based on \cite[Lemma 5]{astolfi2016integral}. Hence, we compare the perturbed system \eqref{eq:CLPH_approx} with the nominal system \begin{equation}\label{eqn:nominal_sys}
    \dot x = \tilde f_d(x),
\end{equation} 
and we aim to study the transfer of stability properties between these two systems under bounded errors in the norm of the model difference. First, notice that by the structural constraints (i)-(iii)  in Proposition~\ref{prop:vanishing_mu}, the equilibrium point $x^\star$ is asymptotically stable on $\cX$ for system \eqref{eqn:nominal_sys}. Moreover, it is also locally exponentially stable. In other words,
the function 
\begin{equation} \label{eqn:local_lyap}
    V(x) = (x-x^\star)^\top \dfrac{\partial^2 \tilde H_d}{\partial x^2}(x^\star) (x-x^\star)
\end{equation}
is a local quadratic Lyapunov function for \eqref{eqn:nominal_sys}.
Indeed, consider the linearized system around $x^\star$ following the dynamics
\begin{equation}\label{eqn:lin_sys}
    \dot {\delta_x}  =  (J-R)\partial H \delta_x,
\end{equation}
where we defined  $\delta_x := x-x^\star, J := \tilde J_d(x^\star)\in \skewsymm{n}, R :=  \tilde R_d(x^\star)\in\pdef{n}, \partial  H : = \tfrac{\partial^2 H_d}{\partial x^2}(x^\star)\in\pdef{n}$, and we exploited the structural constraints \eqref{eq:hamiltonian}. Then, the time derivative of \eqref{eqn:local_lyap} satisfies
\begin{align*}
    \dot V(x) &= \delta_x^\top \partial H \left(J - R + J^\top - R^\top \right)\partial H \delta_x\\
    & = -2\delta_x^\top \partial H R \, \partial H \delta_x < 0,
\end{align*}
where we exploited skew-symmetry and positive definiteness of $J$ and $R$, respectively.
Hence, system \eqref{eqn:lin_sys} is stable and $x^\star$ is a locally exponentially stable equilibrium for \eqref{eqn:nominal_sys}.
Moreover, for any $\varepsilon>0$, the set $\cH_\varepsilon$ contains $x^\star$. Thus, there exist strictly positive scalars $a,b$ such that, for all $x\in \cV_b$, it holds that
\begin{align*}
    \dot V(x) \leq -a V(x).
\end{align*}
We now exploit the locally stable behavior of \eqref{eqn:nominal_sys} to deduce conditions for a locally stable behavior of \eqref{eq:CLPH_approx}.
By \cite[Lemma 5]{astolfi2016integral},
there exists a constant $c>0$ such that, if 
\begin{align}
    \label{eq:xi_bound_1}\left\|\xi(x)\right\|&\le c, \qquad \forall x\in \cH_\varepsilon\\
    \label{eq:xi_bound_2}\left\|\dfrac{\partial \xi(x)}{\partial x}\right\|&\le c, \qquad \forall x\in \cH_\varepsilon,
\end{align}
then system \eqref{eq:CLPH_approx} admits a locally exponentially stable equilibrium point $\bar x$ inside
$\cH_\varepsilon$ with domain of attraction including $\cH_\varepsilon$. Thus, we now aim at deriving conditions on $\mu$ guaranteeing the satisfaction of \eqref{eq:xi_bound_1} and \eqref{eq:xi_bound_2}.
Consider \eqref{eq:xi_bound_1} first. It can be easily verified that, if
\begin{align*}
   \bar \mu &\le  {c}{ \left( \sup_{x\in \cH_\varepsilon}\left\|G^{-1}(x)\right\|\right)^{-1}}\, 
\end{align*}
inequality \eqref{eq:xi_bound_1} is always satisfied.
Consider now \eqref{eq:xi_bound_2}. By exploiting the boundedness of $\mu$, we have
\begin{align*}
    \left\|\dfrac{\partial \xi(x)}{\partial x}\right\|&\le \left\|\dfrac{\partial G^{-1}}{\partial x}(x)\right\|\bar \mu + \left\|G^{-1}(x)\right\|\left\|\dfrac{\partial \mu(x)}{\partial x}\right\|.
\end{align*}
Consequently,
bound \eqref{eq:xi_bound_2} imposes
\[
0\le \left\|\dfrac{\partial \mu(x)}{\partial x}\right\|\leq \dfrac{1}{ \displaystyle \sup_{x\in \cH_\varepsilon}\left\|G^{-1}(x)\right\|}\left(c - \displaystyle\sup_{x\in \cH_\varepsilon}\left\|\dfrac{\partial G^{-1}}{\partial x}(x)\right\|\bar \mu \right),
\]\vspace{-0.5em}
which is well-defined if  $
\bar \mu\le {c}{\Big(  \sup_{x\in \cH_\varepsilon}\left\|\dfrac{\partial G^{-1}}{\partial x}(x)\right\|}\Big)^{-1}\!\!.
$
Then, by defining
\begin{equation*} 
\begin{array}{l}
     c_1 := {c} \left(\max\left({ \displaystyle \sup_{x\in \cH_\varepsilon}\left\|G^{-1}(x)\right\|}, \;  { \displaystyle \sup_{x\in\cH_\varepsilon}\left\|\dfrac{\partial G^{-1}(x)}{\partial x}\right\|}\right)\right)^{-1}, \\
      c_2 := \dfrac{1}{\displaystyle \sup_{x\in \cH_\varepsilon}\left\|G^{-1}(x)\right\|}\left(c - \displaystyle\sup_{x\in \cH_\varepsilon}\left\|\dfrac{\partial G^{-1}(x)}{\partial x}\right\|c_1\right),
\end{array}
\label{eq:total_stab_constr}
\end{equation*}
and selecting $0\le \bar \mu^\star \le \min(c_1,c_2)$, the bounds \eqref{eq:xi_bound_1} and \eqref{eq:xi_bound_2} are always satisfied, thus ensuring existence and local stability of an equilibrium point $\bar x \in \cH_\varepsilon$ for system \eqref{eq:CLPH_approx} by \cite[Lemma 5]{astolfi2016integral}. 
Then, the result follows by noting that, similarly to Proposition~\ref{prop:match_ineq_and_dissip_relax}, inequality  \eqref{eq:damping_assign_relax_tilde} ensures the convergence of the system's trajectories to $\cH_\varepsilon$, which is included in the domain of attraction of $\bar x$.
\end{proof}
Theorem \ref{thm:asym_stab_shifted_eq} establishes local asymptotic stability of $\bar x$ through a partitioning of $\cX$. Trajectories initialized far from $\bar x$ start in the donut-like set $\cD$. Here, the matrix $\tilde R_d(x)$ ensures convergence to the inner boundary, which is shared with the set $\cH_\varepsilon$. Once in $\cH_\varepsilon$, the upper bound constraint on $\mu(x)$
and its Jacobian guarantee the exponential stability of $\bar x$ within $\cH_\varepsilon$.

\begin{remark}
Notice that the structural constraints (i)-(iii) can be relaxed by assuming the existence of a solution $\bT = (J_d,R_d,H_d)$ to the matching equations \eqref{eq:match} that satisfies the assumptions of Proposition~\ref{proposition:idapbc}. This stricter assumption (compared to Standing assumption~\ref{as:approx_match}) only requires the approximations $\tilde J_d, \tilde  R_d, \tilde  H_d$ to satisfy
\[
\left\|f_d(x)-\tilde f_d(x)\right\| \leq \eta, \quad \left\| \frac{\partial f_d}{\partial x}(x)-\frac{\partial \tilde f_d}{\partial x}(x)\right\| \leq \eta, \quad \forall x\in\cX
\]
for some small $\eta\ge 0$. Indeed, by following the same proof of Theorem~\ref{thm:asym_stab_shifted_eq} yet comparing the system to the nominal model \eqref{eq:CLPH}, similar results could be obtained by embedding $\eta$ into $\xi(x)$, namely, by considering the closed loop
\[
\dot x = f_d + \eta + \xi(x) = f_d + \bar \xi(x).
\]
This results in stricter bounds on the NN approximations, but it allows for the use of unstructured approximations. However, from a practical perspective, imposing structural constraints on the NN is often easier than verifying the existence of $\bT$.
\end{remark}

In the next section, we will show how to design neural networks satisfying the required structural constraints by design.

\section{The neural networks design}\label{sec:implementationdetails}
In Section III, we explored the consequences of allowing an approximation error in solving the optimization problem \eqref{eqn:neuralIDAopt}, specifically under compliance with structural constraints (i)-(iii). However, the approach for implementing these approximators was previously unaddressed. We employ NNs to represent the missing functions, leveraging their ability to adjust their weights to meet the established criteria. The focus in this section is to define $\Tilde{J}_d, \Tilde{R}_d, \Tilde{H}_d$ using NNs, ensuring that properties (i)-(iii) are fulfilled, irrespective of their weight's values.
Unlike the unified NN approach presented in \cite{sanchez2021total, sanchez2022under}, we divide the inverse problem, assigning each function a separate NN. Our separation offers three main advantages: (a) the NNs maintain the physical interpretation of the functions they represent, (b) the NNs can be enforced to inherently satisfy boundary conditions and, as a consequence, (c) the optimization process focuses solely in the space of functions that fit within the port-Hamiltonian formulation.

\subsection{NN architectures of $\tilde J_d, \tilde R_d, \tilde H_d$}

To arrive at the desired port-Hamiltonian structure \eqref{eq:CLPH_approx}, we formulate the following NN architectures  for $(\Tilde{J}_d, \Tilde{R}_d, \Tilde{H}_d)$. 
\smallskip

\noindent
{\bfseries Approximating $\mathbf {\tilde J_d}$--} The interconnection function $\Tilde{J}_d: \mathbb{R}^n \rightarrow \skewsymm{n}$ is a skew-symmetric matrix-valued function that describes lossless internal energy exchanges of the target systems. Due to the structure of skew-symmetric matrices, the number of independent terms in $\tilde{J}_d(x)$ is $N_{\text{sk}} =\frac{n(n-1)}{2}$. Hence, by defining a vector of parameters $\theta_j\in\RR^{p_j}$, $\tilde{J}_d$ can be modeled using neural network $\mathrm{NN}_J^{\theta_j}: \RR^{p_j}\times \mathbb{R}^n \rightarrow \mathbb{R}^{N_{\text{sk}}}$ via  
$
\tilde J_d^{\theta_j}(x) = \skewsymmfn{n}(\mathrm{NN}_J^{\theta_j}(x)), 
$
where $\skewsymmfn{n}$ refers to the square skew-symmetric matrix reconstruction map.
\smallskip

\noindent
{\bfseries Approximating $\mathbf {\Tilde{R}_d}$--}  The damping surrogate matrix-valued function $\tilde R_d: \mathbb{R}^n \rightarrow \pdef{n}$, is a function that represents the system's energy dissipation. Notice that, in order to ensure convergence to the equilibrium, we need to restrict the set of allowed dissipation functions to the set of \emph{strictly} positive-definite matrix-valued functions. Hence, $\tilde R_d(x)$ is designed to be     symmetric and positive definite for all $x\in \cD$. The matrix  $\tilde R_d(x)$ comprises of $N_{\text{sy}}=\frac{n(n+1)}{2}$ independent entries. By defining a vector of parameters $\theta_r\in\RR^{p_r}$,  $\tilde{R}_d$ can be modeled by a neural network $\mathrm{NN}_R^{\theta_r}: \RR^{p_r} \times \mathbb{R}^n \rightarrow \mathbb{R}^{N_{\text{sy}}}$ as follows: 
$\tilde{R}_d^{\theta_r}(x) = \symmfn{n}(\mathrm{NN}_R^{\theta_r}(x))+\rho I$,
where $\symmfn{n}$ refers to the square symmetric matrix reconstruction map and
$\rho > |\inf_{x\in\cD} \sigma_{\min}(\symmfn{n}(\mathrm{NN}_R^{\theta_r}(x)))|$ is a positive constant that ensures the positive definiteness of $\tilde R_d$.
\smallskip

\noindent
{\bfseries Approximating $\mathbf {\Tilde{H}_d}$--} Recall that the Hamiltonian surrogate function $\Tilde{H}_d:\RR^n\rightarrow\RR_{\ge0}$ must satisfy the constraints \eqref{eq:hamiltonian}. To enforce them, we adopt the Input-Convex Neural Network (ICNN) framework \cite{kolter2017ICNN}, which guarantees the networks maintain convexity over their input domain $x$ via a careful selection of monotonically increasing activation functions. By selecting a vector of parameters $\theta_h\in\RR^{p_h}$, we first define an intermediate function  $\hat {H}_d$ modeled by a neural network $\mathrm{ICNN}^{\theta_h}: \RR^{p_h} \times \mathbb{R}^n \rightarrow \mathbb{R}$ as follows: 
$\hat{H}_d^{\theta_h}(x) = \mathrm{ICNN}^{\theta_h}(x)$. 
However, to satisfy all the constraints \eqref{eq:hamiltonian}, it is essential to ensure positiveness of $\Tilde{H}_d^{\theta_h}(x)$ and that its minimum coincides with $x^\star$.  Hence, we design $\tilde H_d^{\theta_h}$ as    \begin{equation}\label{eqn:tilde_H_NN}
    \tilde H_d^{\theta_h}(x) = \hat{H}_d^{\theta_h}( \delta_x + \hat{x}) - \hat{H}_d^{\theta_h}(\hat{x}),
\end{equation}  
where $\delta_x  = x- x^\star, \,\hat{x} = \argmin_{x\in\cX} \hat{H}_d(x)$. First, notice that, for any $x\in\RR^n$, the difference in \eqref{eqn:tilde_H_NN} shifts the function's minimum to $0$, due to the definition of $\hat x$. Moreover,  if $x=x^\star$, then $\delta_x = 0$ and $\tilde H_d^{\theta_h}(x)$ = 0. Hence, the minimum of $\tilde H_d^{\theta_h}$ coincides with $x^\star$ and $\tilde {H}_d > 0$ for all $x \in \cX \setminus {x}^\star$. 

\subsection{Loss function formulation}
We now present the training objective. In particular, we present the design of the loss function that follows the approach in \cite{raissi2019PINN}, where the various IDA-PBC design constraints are encoded in the form of residuals.
\smallskip

\noindent
{\bf Matching inequality residual--} This residual corresponds to the matching equation and consists of the minimization of the mismatch function \eqref{eqn:mismatch}. In particular, it describes the local behavior of the desired system.
To set an upper-bound to the mismatch residual, we consider the following relaxed mismatch residual function:
\begin{equation}
\label{eqn:switch_var}
f_m:=(1-d)\|\mu(x)\|, \;
 d := \left\{\begin{array}{l}
      1, \quad \textrm{if }  x\in \cD\cup\partial \cH_\varepsilon,\\
      0, \quad \textrm{otherwise.}
\end{array}\right.
\end{equation}
Note that the condition $x\in \cD\cup\partial \cH_\varepsilon$ can be easily verified by checking  $\tilde H_d^{\theta_h}(x)\ge \varepsilon$, with $\varepsilon>0$. We highlight that  $\varepsilon$ can also be selected as an optimization variable. This residual function rewards the reduction of the mismatch error inside $\cH_\varepsilon$, which affects the distance of the closed-loop equilibrium point from the desired one and the size of its domain of attraction.
\smallskip

\noindent\textbf{Robustness residuals--}
This residual works in synergy with the previous one by describing the desired behavior outside the domain of local stability $\cH_\varepsilon$. More specifically, it aims at guaranteeing convergence to such a set by imposing sufficient dissipation as in \eqref{eq:damping_assign_relax_tilde}. The residual is encoded as the function 
    \begin{align*}
        f_{sd1}&:=d\cdot\max\left(0, \|\xi(x)\| - \kappa \left\|\dfrac{\partial \tilde H_d^{\theta_h}}{\partial x}(x)\right\|\right),\\
        f_{sd2}&:=d\cdot\max(0,\kappa - \sigma_{\min}(\tilde R_d^{\theta_r}(x))),
    \end{align*}
with $\kappa \in \left(0,\rho\right)$ and $d$ the switching variable defined in \eqref{eqn:switch_var}. 
\smallskip

\noindent\textbf{Local exponential stability residual--} According to Theorem~\ref{thm:asym_stab_shifted_eq}, sole minimization of the mismatch error may not be sufficient to achieve local exponential stability. Hence, as suggested by \eqref{eqn:mu_bound_les}, we introduce the additional residual 
\[
f_{dm}:=(1-d)\left\|\frac{\partial \mu}{\partial x}(x)\right\|.
\]
Paired with \eqref{eqn:switch_var}, it encourages satisfaction of the local conditions in Theorem~\ref{thm:asym_stab_shifted_eq}. We remark that such a derivative can be easily computed with automatic differentiation tools.

Our approach integrates the IDA-PBC into a loss function composed of multiple terms, thus raising concerns about their balance, a known issue in PINNs' training \cite{krishnapriyan2021difficultPINN, basir2022difficultPINN}. To overcome these challenges, we adopt a novel strategy as introduced in \cite{anagnostopoulos2023residualPINN}, which involves applying an adaptive scaling factor to each objective term. This scaling factor is adjusted during the optimization process, enabling the optimizer to even out the impact of each individual term during the initial optimization phases. This also helps the optimization algorithm to focus on the most significant penalties at each iteration, promoting a more effective and balanced process.
Using the adaptive scaling factor strategy, the final loss function is given by $\cL:= r_m f_m + r_{dm} f_{dm} + r_{sd} (f_{sd1}+f_{sd2})$, and the update rule that scales multipliers is defined as $$r_i(t+1) = \gamma r_i(t) + \frac{\alpha f_i}{\max_{i}(f_i)}, $$ for $i \in \{m, dm, sd\}$, where the decay rate $\gamma \in (0,1)$ and update rate $\alpha >0$  are 
design parameters.

\section{Simulations}\label{sec:simulations}
In this section, we test the proposed methodology on three academic examples: a fully-actuated double pendulum system, an under-actuated cartpole system, and a nonlinear mass-spring-damper system. In all cases, assuming the dissipation condition \eqref{eq:damping_assign_relax_tilde} holds, the closed-loop is guaranteed to present an asymptotically stable equilibrium point by Theorem~\ref{thm:asym_stab_shifted_eq} if the local approximation of the matching inequalities is sufficiently precise. Otherwise, Proposition~\ref{prop:match_ineq_and_dissip_relax} guarantees practical stability of $x^\star$. 
In the first example, the matching equation can be solved explicitly and the use of \emph{Neural IDA-PBC} enables the automated discovery of the IDA-PBC solution. For the second one, to the best of the authors' knowledge, there is no exact solution to the matching equations due to the necessity of \emph{kinetic-energy shaping}, and neural IDA-PBC allows designing a guaranteed controller under approximate matching conditions. In the third case, \emph{Neural IDA-PBC} is applied to a system modeled in a general input affine form, showcasing the independence from the modeling framework. In all cases, the training data set is randomly sampled at the beginning of each epoch to ensure a thorough exploration of the training space. The training hyperparameters are presented in Table~\ref{tab:optimization}. The NN parameters are trained with  ADAM \cite{Kingma2014AdamAM}.

\subsection{Benchmarks}
For the double pendulum and cartpole examples, the open-loop interconnection and dissipation matrices are defined by
\begin{equation*}
    J(x)= \begin{bmatrix}
        0^{n/2\times n/2} &I^{n/2\times n/2}\\ -I^{n/2\times n/2} &0^{n\times n}
    \end{bmatrix},\quad
    R(x)=0^{n\times n}
\end{equation*}
and the Hamiltonian is $\cH(x) = \frac{1}{2}p^\top M(x)^{-1} p + U(x),$ with $x=(q,p)$, where $q$ and $p$ are the generalized coordinates and momenta respectively. $M(x)$ and $U(x)$ represent the inertia and potential energy functions and will be defined later.
\smallskip

\noindent
{\textbf{Double Pendulum--}}
A fully-actuated double pendulum has to be stabilized in
the vertical position. The system's open-loop dynamics \eqref{eq:OLPH} are fully defined by the functions
\begin{align*}
    M(x) &= \begin{bmatrix}
(m_1+m_2)l_1^2 & \frac{1}{2}m_2 l_1 l_2 \cos(q_1-q_2) \\
\frac{1}{2}m_2 l_1 l_2 \cos(q_1-q_2) & m_2 l_2^2
 \end{bmatrix},\\
 U(x) &= m_2 g (l_1 (1+\cos(q_1)) + l_2 (1+\cos(q_2)))\\
 &\qquad +  m_1 g l_1 (1+\cos(q_1)), 
\end{align*}
where $m_i,l_i$ are the mass and length of link $i$, respectively.
In our simulations, $m_1=m_2=l_1=l_2=1$.
\smallskip

\noindent
\textbf{Cartpole--}
An inverted pendulum on a cart sliding on a rail  (cartpole) has to be stabilized in the vertical position, while the cart must return to the origin. The system's input is the force applied to the cart. By defining $m_1$ as the mass of the cart and $m_2,l_2$ as the mass and length of the pendulum respectively, the system's open-loop dynamics \eqref{eq:OLPH} are fully defined by
\begin{align*}
    M(x) &= \begin{bmatrix}
m_1+m_2 & \frac{1}{2}m_2 l_2 \cos(q_2) \\
\frac{1}{2}m_2 l_2 \cos(q_2) & m_2 l_2^2
 \end{bmatrix},\\
 U(x) &= m_2 g l_2 (1+\cos(q_2) ).
\end{align*}
In our simulations, we set $m_1=m_2=l_2=1$. 
\smallskip

\noindent\textbf{Nonlinear-Mass-spring-damper--}
A nonlinear-mass-spring-damper system with a softening spring force: $F_k = q(k + 0.5q^2)$ and a nonlinear damping force: $F_d = bp^2(1 + |p|)$ has to be stabilized at the arbitrary position $x^\star=(0.5, 0.0)$. The system is
modeled in a general non-port-Hamiltonian framework and evolves according to the dynamics $\dot x = f(x)+g(x)u$ with
\begin{equation*}
    f(x)= Ax = \begin{pmatrix}
    0 & 1\\
    -\frac{(k+0.5q^2)}{m} & -\frac{bp(1 + |p|)}{m}
    \end{pmatrix} x, \quad g(x) =
    \begin{pmatrix}
    0\\
    1
    \end{pmatrix},
\end{equation*}
with $m=k=1$ and $b=0.5$.
\begin{table}[!t]
\caption{Optimization details}\label{tab:optimization}
\centering
\begin{tabular}{ cc}
 Parameter & Value\\
 \hline
 $H_d$ NN size & $n\times3\times300\times1$\\
 $J_d$ NN size & $n\times3\times300\times N_{sk}$\\
 $R_d$ NN size & $n\times3\times300\times N_{sy}$\\
 $\#$ epochs   & 2000\\
 $\#$ samples/epoch & 512\\
 $\gamma$ & 0.9999\\
 $\alpha$ & 0.01\\
\hline
\end{tabular}
\end{table}

\subsection{Results}

\begin{figure*}
    \centering
    \subfloat[Double pendulum\label{fig:doublependulum}]{\includegraphics[width=0.716\textwidth]{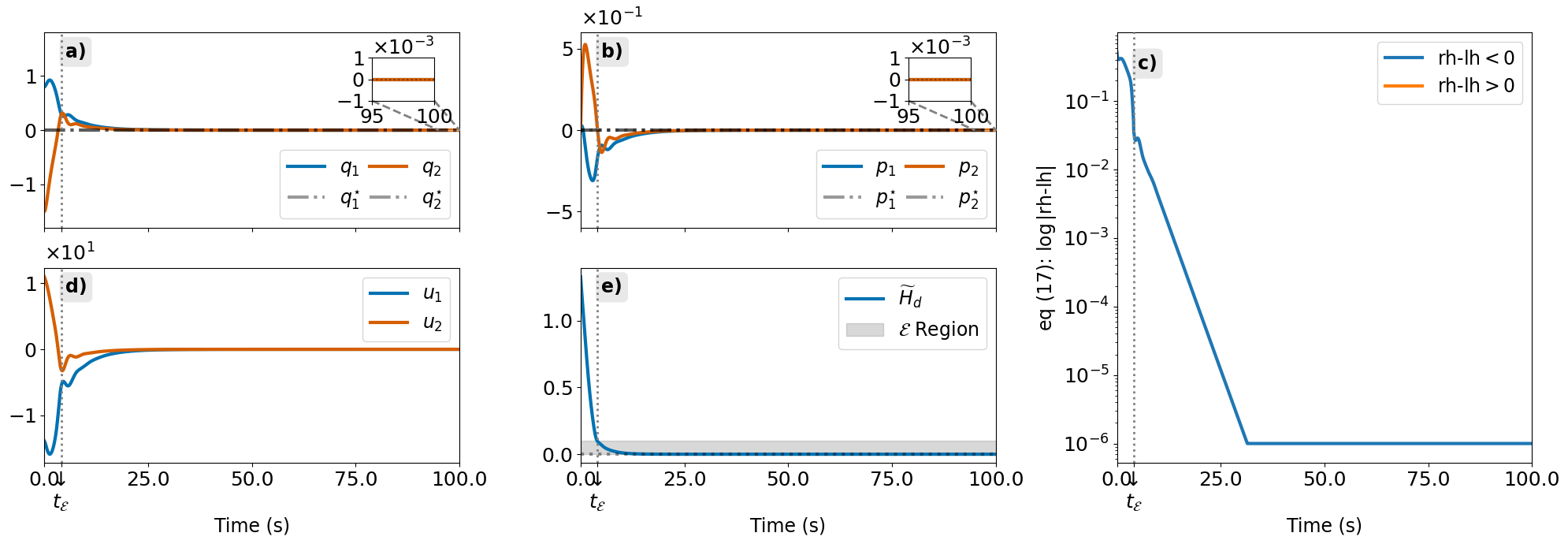}}\\
    \subfloat[Double pendulum with extra damping\label{fig:doublependulumdamped}]{\includegraphics[width=0.716\textwidth]{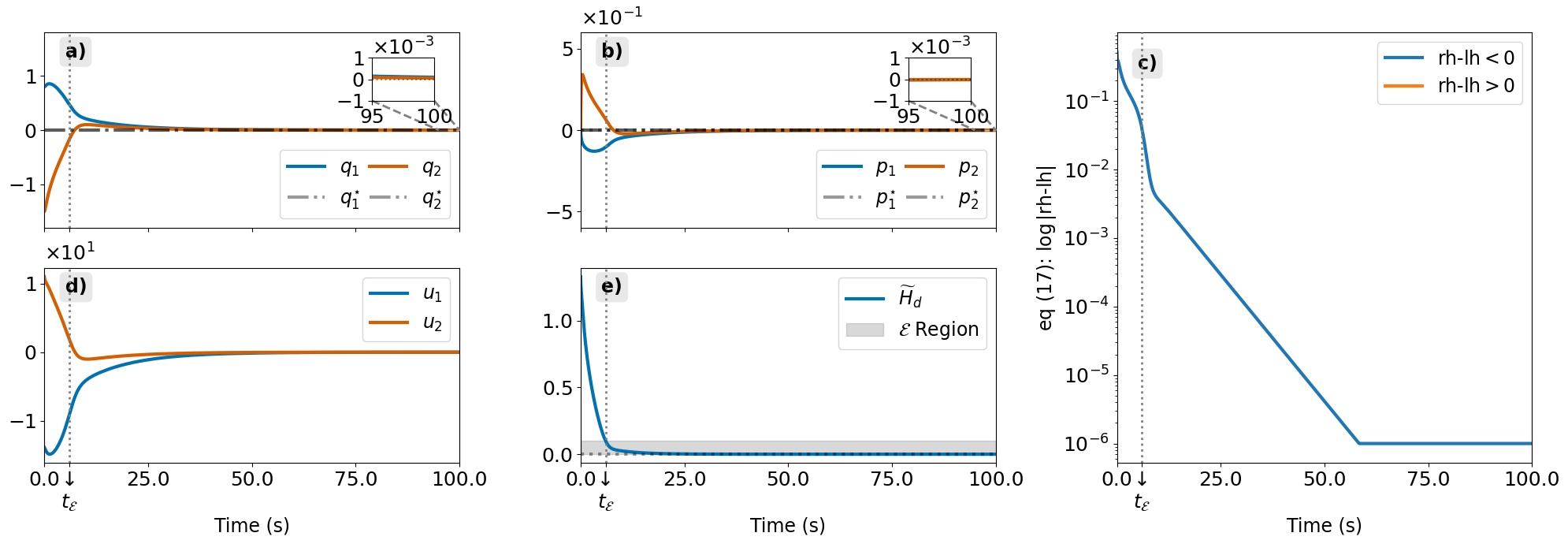}}\\
    \subfloat[Cartpole\label{fig:cartpole}]{\includegraphics[width=0.716\textwidth]{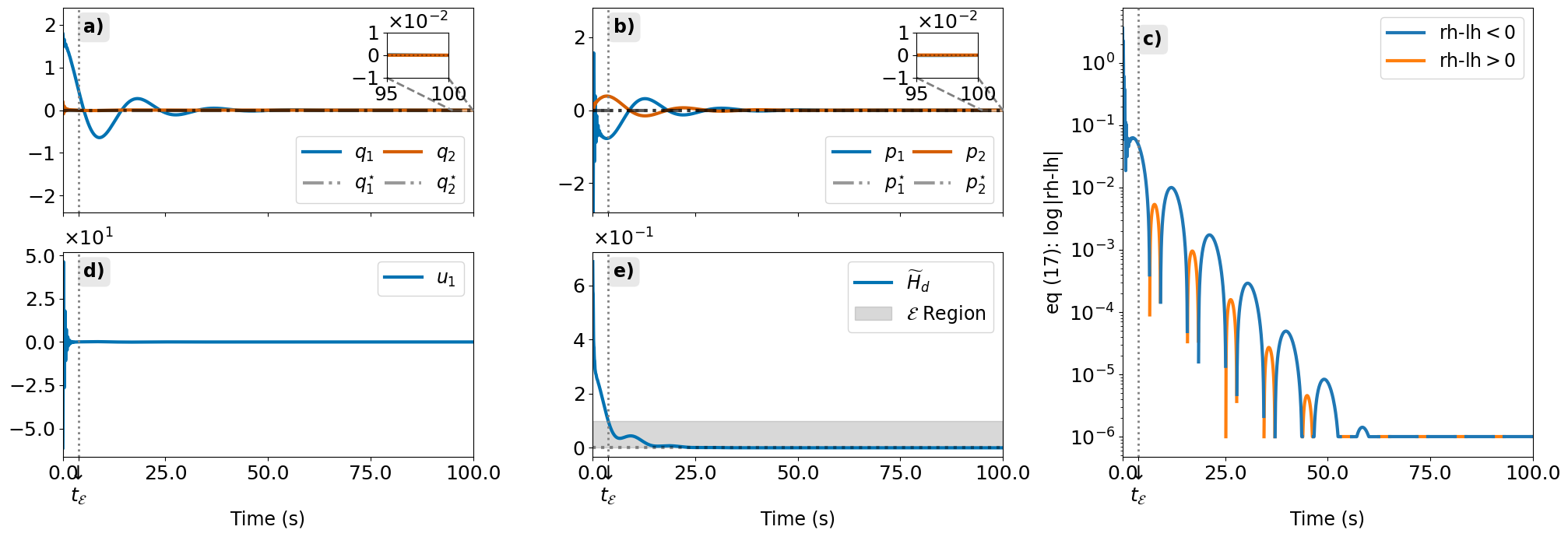}}\\
    \subfloat[Nonlinear-mass-spring-damper    \label{fig:massspringdamper}]{\includegraphics[width=0.716\textwidth]{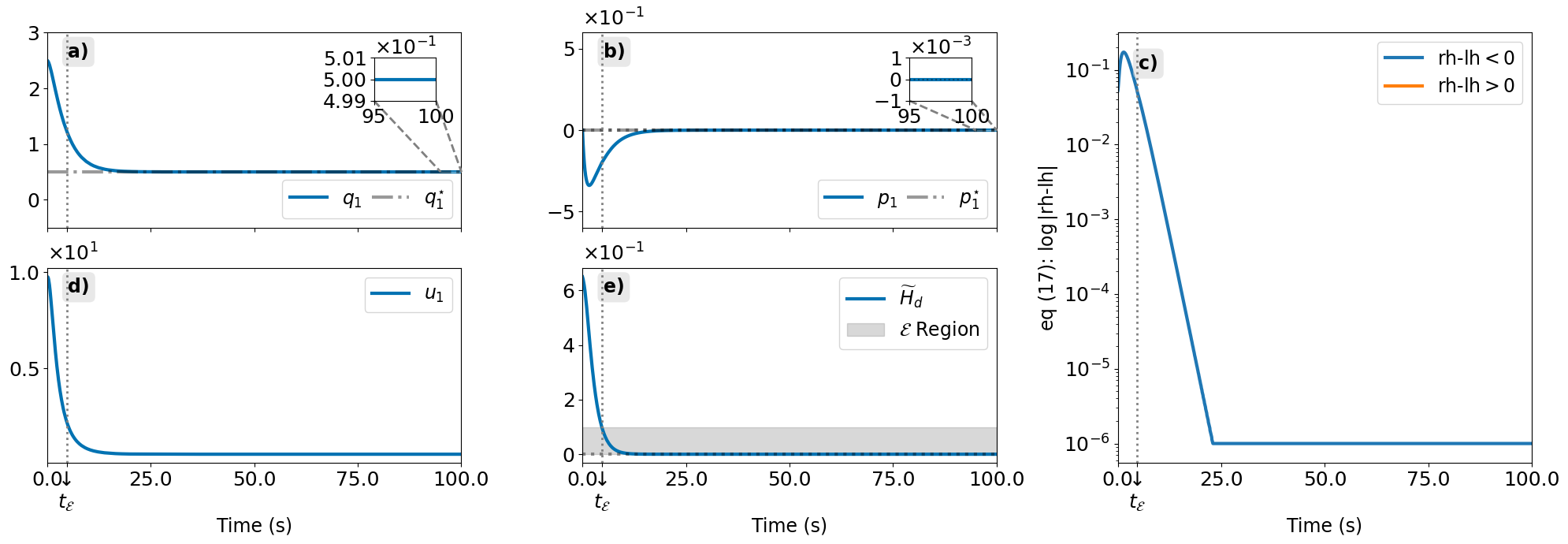}}
    \caption{Simulation results for the different benchmarks: a) and b) are the generalized coordinates and momenta; c) depicts the satisfaction of condition \eqref{eq:damping_assign_relax_tilde}; d) and e) are the \emph{Neural IDA-PBC} control signal and closed-loop energy. The gray area in e) determines the region $\cE$ where the sufficient dissipation condition is not necessarily satisfied.
    The vertical (dotted) line indicates the time at which the trajectories enter $\cE$.}
    \label{fig:simulationdynamics}
\end{figure*}

Figure \ref{fig:simulationdynamics} presents the trajectories of relevant quantities for the three benchmarks. 
Plots a) and b),  show the system's trajectories initiated at arbitrary initial conditions. For all benchmarks, the system is stabilized to an equilibrium point.
Plots c) present the difference between the right-hand side and the left-hand side of \eqref{eq:damping_assign_relax_tilde}. All plots have been clamped once reaching a threshold of $10^{-6}$ to improve readability. It can be seen that all benchmarks are at a \emph{guaranteed} sufficient dissipation regime before entering $\cE$, being $\cE$ the \emph{estimated} set of local exponential stability $\cE ={\cH}_{\widetilde\varepsilon}$, with $\tilde \varepsilon$ the estimated value for $\varepsilon$. We identify this entering time instant by $t_{\cE}$. Even if most examples achieve sufficient dissipation on the whole domain, inside $\cE$ the sufficient dissipation condition is not necessarily satisfied. Indeed, if an exact solution exists, the local minimization of $\|\mu(x)\|$ can lead to such a condition being verified up to a negligibly small region around the desired equilibrium. Nevertheless, if such a condition is not verified,  trajectories still converge to a fixed point $\bar x$ which is sufficiently close to the target $x^\star$, as shown in (iii.c). 
This is in accordance with the results of Theorem~\ref{thm:asym_stab_shifted_eq}, indicating that the matching conditions are locally well-approximated. Graph d) shows the evolution of the control signal resulting from the approximated solutions. Graph e) shows that the approximated $\tilde{H}_d$ is always decreasing whenever \eqref{eq:damping_assign_relax_tilde} is satisfied. 
Interestingly, in (iii.c) $\tilde H_d$ increases and it is not a valid Lyapunov function inside $\cE$.  As before, this is in accordance with the proof of Theorem~\ref{thm:asym_stab_shifted_eq}, where local stability is proven with a quadratic function not necessarily matching $\tilde H_d$. Moreover, this strengthens the intuition that the system is stabilized in  a small neighborhood of $x^\star$.
Finally, plots (ii) present the results under a control law where $\rho$ in $\tilde{R}_d^{\theta_r}(x)$ has been modified post-training. This showcases another advantage of our modular approach, as it allows fine-tuning of the transient behavior without the need of retraining.

\section{Conclusions}\label{sec:conclusions}
We studied and restructured \emph{Neural IDA-PBC}, a methodology employing NNs in the design of IDA-PBC control. First, we analyzed the closed-loop system under approximate solutions to the matching equation. By interpreting the approximation errors as external additive perturbations and exploiting the nominal solution's robustness properties, we proved asymptotic stability of the desired equilibrium under the NN-based solution. 
Second, we proposed a novel implementation of \emph{Neural IDA-PBC} and demonstrated its effectiveness on three benchmarks. 
Our renewed methodology enables physics-based feedback control for input-affine nonlinear systems, avoiding the need for \emph{ad-hoc} designs. The use of multiple NNs allows the preservation of the fundamental physical interpretation of port-Hamiltonian dynamics. Moreover, this separation allows for post-training fine-tuning of transient behaviors. 
The optimization-centric nature of \emph{Neural IDA-PBC} allows the incorporation of performance metrics into the optimization problem. Future works will focus on including those metrics in the optimization objective.

\begin{ack}                               
We would like to acknowledge R. Reyes-Baez for the stimulating discussion and early works on Neural IDA-PBC method \cite{sanchez2021total,sanchez2022under}. This publication is part of the project Digital Twin project 6 with project number P18-03 of the research programme Perspectief which is (mainly) financed by the Dutch Research Council (NWO).
\end{ack}

\bibliographystyle{abbrv}        
\bibliography{autosam}           
\end{document}